\documentclass[12pt]{article}%
\usepackage{amsmath,latexsym}
\usepackage{graphicx}
\usepackage{amsmath}
\usepackage{amsfonts}
\usepackage{amssymb}%
\begin{document}

\title{{Wormholes with a barotropic equation of state
   admitting a one-parameter group of conformal motions}}
   \author{
  Peter K.F. Kuhfittig\\  \footnote{kuhfitti@msoe.edu}
 \small Department of Mathematics, Milwaukee School of
Engineering,\\
\small Milwaukee, Wisconsin 53202-3109, USA}

\date{}
 \maketitle
\noindent\textbf{Highlights}\\
The theoretical construction of Morris-Thorne wormholes
retains complete control over the geometry at the expense
of the stress-energy tensor.\\
The introduction of a barotropic equation of state fails to
produce a solution even if the energy density is known.\\
The assumption of conformal symmetry fills the gap in the
form of a complete wormhole solution.\\

\begin{abstract}\noindent
The theoretical construction of a traversable
wormhole proposed by Morris and Thorne maintains
complete control over the geometry by assigning
both the shape and redshift functions, thereby
leaving open the determination of the
stress-energy tensor.  This paper examines the
effect of introducing the linear barotropic
equation of state $p_r=\omega\rho$ on the
theoretical construction.  If either the
energy density or the closely related shape
function is known, then the Einstein field
equations do not ordinarily yield a finite
redshift function.  If, however, the wormhole
admits a one-parameter group of conformal
motions, then both the redshift and shape
functions exist provided that
$\omega <-1$.  In a cosmological setting,
the equation of state $p=\omega\rho$,
$\omega <-1$, is associated with phantom
dark energy, which is known to support
traversable wormholes.    \\

\end{abstract}

\section{Introduction}\label{E:introduction}

Wormholes are tunnel-like structures connecting
different universes or widely separated regions
of our own Universe.  That wormholes could be
actual physical objects was first proposed by
Morris and Thorne \cite{MT88}, who assumed that
the wormhole spacetime can be described by the
following static spherically symmetric line
element:
\begin{equation}\label{E:line1}
ds^{2}=-e^{2\Phi(r)}dt^{2}+\frac{dr^2}{1-b(r)/r}
+r^{2}(d\theta^{2}+\text{sin}^{2}\theta\,
d\phi^{2}).
\end{equation}
The function $\Phi=\Phi(r)$ is called the
\emph{redshift function}, which must be
everywhere finite to prevent an event horizon.
The function $b=b(r)$ is called the \emph{shape
function} because it determines the spatial
shape of the wormhole when viewed, for example,
in an embedding diagram.  The spherical surface
$r=r_0$ is the \emph{throat} of the wormhole
and must satisfy the following conditions:
$b(r_0)=r_0$, $b(r)<r$ for $r>r_0$, and
$b'(r_0)<1$, usually referred to as the
\emph{flare-out condition}.  This condition
refers to the flaring out of the embedding
diagram pictured in Ref. \cite{MT88}.  The
flare-out condition can only be met by
violating the null energy condition.

Using units in which $c=G=1$,
the Einstein field equations in the orthonormal
frame, $G_{\hat{\mu}\hat{\nu}}=8\pi
T_{\hat{\mu}\hat{\nu}}$, yield the following
simple interpretation for the components of the
stress-energy tensor: $T_{\hat{t}\hat{t}}=
\rho(r)$, the energy density,
$T_{\hat{r}\hat{r}}=p_r$, the radial pressure,
and $T_{\hat{\theta}\hat{\theta}}=
T_{\hat{\phi}\hat{\phi}}=p_t$, the lateral
pressure.  For the theoretical construction of
the wormhole, Morris and Thorne then proposed
the following strategy: retain complete control
over the geometry by specifying the functions
$b(r)$ and $\Phi(r)$ to obtain the desired
properties.  It is then up to the engineering
team to search for or to manufacture the
materials or fields that yield the required
stress-energy tensor.

Researchers have tried various strategies for
meeting this goal, some of which are discussed
in the next section.  These include the main
subjects of this paper, the introduction of
an appropriate equation of state and the
geometric assumption of conformal symmetry,
which leads to motions for which the metric
tensor is invariant up to a scale factor.

\section{Some stategies}

To describe the various strategies, we must
first list the Einstein field equations:
\begin{equation}\label{E:Einsteina}
  \rho(r)=\frac{b'}{8\pi r^2},
\end{equation}
\begin{equation}\label{E:Einsteinb}
   p_r(r)=\frac{1}{8\pi}\left[-\frac{b}{r^3}+
   2\left(1-\frac{b}{r}\right)\frac{\Phi'}{r}
   \right],
\end{equation}
\begin{equation}\label{E:Einsteinc}
   p_t(r)=\frac{1}{8\pi}\left(1-\frac{b}{r}\right)
   \left[\Phi''-\frac{b'r-b}{2r(r-b)}\Phi'
   +(\Phi')^2+\frac{\Phi'}{r}-
   \frac{b'r-b}{2r^2(r-b)}\right].
\end{equation}
Eq. (\ref{E:Einsteinc}) can actually be obtained
from the conservation of the stress-energy tensor,
i.e., $T^{\mu\nu}_{\phantom{\mu\nu};\nu}=0$; so only
two of Eqs. (\ref{E:Einsteina})-(\ref{E:Einsteinc})
are independent.  As a result, one can simply
write
\begin{equation}\label{E:Alberta}
  b'=8\pi\rho r^2
\end{equation}
and
\begin{equation}
  \Phi'=\frac{8\pi p_rr^3+b}{2r(r-b)}.
\end{equation}

Next, suppose we adopt the linear barotropic
equation of state $p=\omega\rho$, which has
been used in various cosmological settings.
Since we are now dealing with wormholes, we will
adopt the form
 \begin{equation}\label{E:EoS}
    p_r=\omega\rho.
\end{equation}
So if $\rho(r)$, or equivalently, $b(r)$ is
assigned, one could conceivably obtain
$p_r(r)$ and $\Phi(r)$ and hence a complete
description of the wormhole geometry.

The energy density may also be known for
physical reasons.  One possibility is the
Navarro-Frenk-White density profile in the
dark-matter halo \cite{NFW, fR13}:
\begin{equation}
  \rho(r)=\frac{\rho_s}{\frac{r}{r_s}
  \left(1+\frac{r}{r_s}\right)^2}.
\end{equation}
Here $r_s$ is the characteristic scale radius
and $\rho_s$ the corresponding density.

Another possible physical reason for having
$\rho(r)$ is noncommutative geometry,
which replaces point particles by smeared
objects in order to eliminate some of the
divergences that normally appear in general
relativity.  The smearing effect is modeled
by the use of either the Gaussian curve of
minimal length $\sqrt{\theta}$ to represent the
energy density \cite{SW99, NSS06, SS09, fR12a, pK13}
\begin{equation}
   \rho(r)=\frac{M}{(4\pi\theta)^{3/2}}
   e^{-r^2/4\theta}
\end{equation}
or by the Lorentzian curve \cite{LL12}
\begin{equation}
    \rho(r)=\frac{M\sqrt{\theta}}{\pi^2(r^2+\theta)^2}.
\end{equation}
Here the mass $M$, instead of being perfectly
localized, is diffused throughout the region due
to the uncertainty.  To clarify this statement,
observe that the mass $M_{\theta}$ inside a
sphere of radius $r$ is
\begin{equation*}
   M_{\theta}=\int_{r_0}^r\rho(r')4\pi (r')^2dr'
   =\frac{2M}{\pi}\left(\text{tan}^{-1}
   \frac{r}{\sqrt{\theta}}-
   \frac{r\sqrt{\theta}}{r^2+\theta}\right);
\end{equation*}
thus $M_{\theta}\rightarrow M$ as $\theta
\rightarrow 0$.  (In particular, when viewed
from a distance, $M_{\theta}=M$.)

For all these cases, $b(r)$ can be computed from
Eq. (\ref{E:Alberta}).  So Eq. (\ref{E:EoS})
then yields
\begin{equation*}
  \frac{1}{8\pi}\frac{b'}{r^2}=\frac{1}{\omega}
  \left(\frac{1}{8\pi}\right)\left[-\frac{b}{r^3}
  +2\left(1-\frac{b}{r}\right)\frac{\Phi'}{r}
  \right]
\end{equation*}
and
\begin{equation}\label{E:Phiprime}
  \Phi'=\frac{\omega rb'+b}{2r^2
  \left(1-\frac{b}{r}\right)}.
\end{equation}

Recalling that $r=r_0$ at the throat, $\Phi'$
and hence $\Phi$ are not likely to be defined
at $r=r_0$, thereby yielding an event horizon
for negative $\omega$, as, for example, in
Ref. \cite{sS05}.  A proper choice of $b(r)$
can avoid this problem:
$b(r)=r_0(r/r_0)^{-1/\omega}$ leads to
$\Phi'\equiv 0$, thereby avoiding an event
horizon \cite{fL06}.

A better way is to obtain $\Phi=\Phi(r)$
independently of the field equations.  Thus
Ref. \cite{Kuhfittig} relies on both
dark-matter and dark-energy models to obtain
$\Phi(r)$.

In this paper we propose another method, the
assumption that the wormhole admits a
one-parameter group of conformal motions,
discussed next.

\section{Conformal Killing vectors}

As noted in the Introduction, we assume that our
spacetime admits a one-parameter group of conformal
motions, which are motions along which the metric
tensor of a spacetime remains invariant up to a
scale factor.  This is equivalent to the existence
of conformal  Killing vectors such that
\begin{equation}\label{E:Lie}
   \mathcal{L_{\xi}}g_{\mu\nu}=g_{\eta\nu}\,\xi^{\eta}
   _{\phantom{A};\mu}+g_{\mu\eta}\,\xi^{\eta}_{\phantom{A};
   \nu}=\psi(r)\,g_{\mu\nu},
\end{equation}
where the left-hand side is the Lie derivative of the
metric tensor and $\psi(r)$ is the conformal factor.
(For further discussion, see \cite{MM96, BHL07}.)
The vector $\xi$ generates the conformal
symmetry and the metric tensor $g_{\mu\nu}$ is
conformally mapped into itself along $\xi$.  This
type of symmetry has been used to great advantage
in describing relativistic stellar-type objects
\cite{HPa, HPb}.  Besides leading to new solutions, the
conformal symmetries have led to new geometric and
kinematical insights \cite{MS93, Ray08, fR10, fR12b}.

Exact solutions of traversable wormholes admitting conformal
motions have also been found, given a noncommutative-geometry
background \cite{R2K3}.  Two earlier studies assumed a
\emph{non-static} conformal symmetry \cite{BHL07, BHL08}.

To discuss conformal symmetry, it is convenient to use
the following form of the metric \cite{R2K3}:
\begin{equation}\label{E:line2}
   ds^2=- e^{\nu(r)} dt^2+e^{\lambda(r)} dr^2
   +r^2( d\theta^2+\text{sin}^2\theta\, d\phi^2).
\end{equation}
Now the Einstein field equations take on the
following form:

\begin{equation}\label{E:Einstein1}
e^{-\lambda}
\left[\frac{\lambda^\prime}{r} - \frac{1}{r^2}
\right]+\frac{1}{r^2}= 8\pi \rho,
\end{equation}

\begin{equation}\label{E:Einstein2}
e^{-\lambda}
\left[\frac{1}{r^2}+\frac{\nu^\prime}{r}\right]-\frac{1}{r^2}=
8\pi p_r,
\end{equation}

\noindent and

\begin{equation}\label{E:Einstein3}
\frac{1}{2} e^{-\lambda} \left[\frac{1}{2}(\nu^\prime)^2+
\nu^{\prime\prime} -\frac{1}{2}\lambda^\prime\nu^\prime +
\frac{1}{r}({\nu^\prime- \lambda^\prime})\right] =8\pi p_t.
\end{equation}

Next, we turn our attention to the assumption of conformal
symmetry in Eq. (\ref{E:Lie}).  Here we follow Herrera and
Ponce de Le\'{o}n \cite{HPa} and restrict the vector field
by requiring that $\xi^{\alpha}U_{\alpha}=0$, where
$U_{\alpha}$ is the four-velocity of the perfect fluid
distribution.  The assumption of spherical symmetry then
implies that $\xi^0=\xi^2=\xi^3=0$ \cite{HPa}.  Eq. (\ref
{E:Lie}) now yields the following results:
\begin{equation}\label{E:sol1}
    \xi^1 \nu^\prime =\psi,
\end{equation}
\begin{equation}\label{E:sol2}
   \xi^1  = \frac{\psi r}{2},
\end{equation}
and
\begin{equation}\label{E:sol3}
  \xi^1 \lambda ^\prime+2\,\xi^1 _{\phantom{1},1}=\psi.
\end{equation}

From these equations we obtain
\begin{equation} \label{E:gtt}
   e^\nu  =C_1 r^2
\end{equation}
and
\begin{equation}\label{E:grr}
   e^\lambda  = \left(\frac {a} {\psi}\right)^2,
\end{equation}
where $C_1$ and $a$ are integration constants.  In
order to make use of Eqs. (\ref{E:gtt}) and
(\ref{E:grr}), we rewrite Eqs.
(\ref{E:Einstein1})-(\ref{E:Einstein3}) as follows:
\begin{equation}\label{E:E1}
\frac{1}{r^2}\left(1 - \frac{\psi^2}{a^2}
\right)-\frac{2\psi\psi^\prime}{a^2r}= 8\pi \rho,
\end{equation}
\begin{equation}\label{E:E2}
\frac{1}{r^2}\left( \frac{3\psi^2}{a^2}-1
\right)= 8\pi p_r,
\end{equation}
and
\begin{equation}\label{E:E3}
\frac{\psi^2}{a^2r^2}
+\frac{2\psi\psi^\prime}{a^2r} =8\pi p_t.
\end{equation}

\section{Wormhole structure}

Returning to the equation of state (\ref{E:EoS}),
$p_r=\omega\rho$, Eqs. (\ref{E:E1}) and (\ref{E:E2})
yield (after some simplification)
\begin{equation}\label{master}
   2r\omega\Psi\Psi'+(\omega +3)\Psi^2=
   a^2(\omega +1).
\end{equation}
Noting that $2\psi\psi'=(\psi^2)'$, the equation
is linear in $\psi^2$ and can be readily solved
to yield
\begin{equation}\label{E:closedform}
    \Psi^2(r)=\frac{1}{\omega+3}
    [a^2(\omega+1)+(\omega+3)
    cr^{-(\omega+3)/\omega}],
\end{equation}
where $c$ is an integration constant.  Comparing
line elements (\ref{E:line1}) and (\ref{E:line2}),
we have in view of Eq. (\ref{E:grr}),
\begin{equation}\label{E:shape1}
  b(r)=r(1-e^{-\lambda})=
      r\left(1-\frac{\psi^2}{a^2}\right),
\end{equation}
which yields the following class of shape
functions:
\begin{equation}\label{E:shape2}
   b(r)=r\left(\frac{2}{\omega+3}
   -\frac{c}{a^2}r^{-(\omega+3)/\omega}
   \right).
\end{equation}
The requirement $b(r_0)=r_0$ can be used to
determine the integration constant.  In particular,
from Eq. (\ref{E:shape2}),
\[
   \frac{2}{\omega+3}-\frac{c}{a^2}
   r_0^{-(\omega+3)/\omega}=1
\]
and
\[
    c=-\frac{\omega+1}{\omega+3}a^2
    r_0^{(\omega+3)/\omega}.
\]
So Eq. (\ref{E:shape2}) becomes
\begin{equation}\label{E:shape3}
   b(r)=r\left(\frac{2}{\omega+3}+
   \frac{\omega+1}{\omega+3}
   r_0^{(\omega+3)/\omega}
       r^{-(\omega+3)/\omega}\right).
\end{equation}

The next step is to check the flare-out condition
$b'(r_0)<1$.  From Eq. (\ref{E:shape3}),
\begin{multline}
   b'(r)=\frac{2}{\omega+3}+
   \frac{\omega+1}{\omega+3}
   r_0^{(\omega+3)/\omega}
   r^{-(\omega+3)/\omega}\\
   +r\left[\frac{\omega+1}{\omega+3}
   r_0^{(\omega+3)/\omega}\left(
   -\frac{\omega+3}{\omega}
   r^{-(\omega+3)/\omega-1}\right)\right].
\end{multline}
After substituting $r_0$ for $r$ and
simplifying, we obtain
\begin{equation}
    b'(r_0)=-\frac{1}{\omega}<1
\end{equation}
provided that $\omega <-1$.  So the flare-out
condition is met.

This result shows that a wormhole solution
requires a phantom-energy background, i.e.,
$\omega <-1$, which is consistent with
earlier studies \cite{sS05, fL05, pK09, pK10}.
(The reason is that whenever $\omega <-1$
in the equation of state $p=\omega\rho$,
the null energy condition is violated.)

To obtain the redshift function from Eq.
(\ref{E:gtt}), we need to determine the
integration constant $C_1$, discussed next.

\section{Junction to an external vacuum solution}

Eq. (\ref{E:gtt}) implies that the wormhole
spacetime is not asymptotically flat.  So the
wormhole material must be cut off at some
$r=r_1$ and joined to the exterior
Schwarzschild solution
\begin{equation}
ds^{2}=-\left(1-\frac{2M}{r}\right)dt^{2}
+\frac{dr^2}{1-2M/r}
+r^{2}(d\theta^{2}+\text{sin}^{2}\theta\,
d\phi^{2}).
\end{equation}
Here
\begin{equation}\label{E:M}
   M=\frac{1}{2}b(r_1)=
   \frac{1}{2}r_1\left(\frac{2}{\omega+3}
   +\frac{\omega+1}{\omega+3}
     r_0^{(\omega+3)/\omega}
        r_1^{-(\omega+3)/\omega}\right).
\end{equation}
So for $e^{\nu}=C_1r^2$ from Eq. (\ref{E:gtt}),
we have $C_1r_1^2=1-2M/r_1$ and the integration
constant becomes
\begin{equation}
  C_1=\frac{1}{r_1^2}\left(1-\frac{2M}{r_1}
  \right),
\end{equation}
where $M$ is given in Eq. (\ref{E:M}).  This
completes the wormhole solution.

\section{Conclusion}

After establishing that the adoption of the
linear barotropic equation of state $p_r=\omega\rho$
is usually insufficient for obtaining a finite
redshift function from the Einstein field equations,
it is shown in this paper that such a redshift
function can be obtained by assuming that the
wormhole admits a one-parameter group of conformal
motions.  The solution requires a phantom-energy
background, however, a conclusion that is consistent
with earlier studies.  The resulting wormhole
is not asymptotically flat and must be joined to
an external Schwarzschild spacetime.

\end{document}